\documentclass[nofootinbib]{revtex4}
\usepackage[english]{babel}
\usepackage{array,booktabs}   
\usepackage{array} 
\usepackage{amsmath} 
\usepackage{relsize}
\usepackage{wrapfig}
\usepackage{calc}
\usepackage{pdflscape}
\usepackage{color}
\usepackage{float}
\usepackage[font=small,labelfont=bf]{caption}
\usepackage{graphicx}
\usepackage{amsmath}
\usepackage[nodisplayskipstretch]{setspace}

\usepackage{setspace}
\usepackage{tabularx}

\usepackage{float}
\usepackage{color}
\usepackage{amsmath}
\usepackage{float}
\usepackage{calc}
\usepackage{pdflscape}
\usepackage{color}
\usepackage{float}
\usepackage{graphics}
\usepackage{epsfig}
\usepackage{epstopdf}
\usepackage{amssymb}
\usepackage[font=small,labelfont=bf]{caption}
\usepackage{graphicx}
\usepackage{epstopdf}
\begin{document}{\setlength\abovedisplayskip{4pt}}

\title{Atmospheric muon charge ratio analysis at the INO-ICAL detector}
\author{Jaydip Singh\footnote{E-mail: jaydip.singh@gmail.com}, Jyotsna Singh\footnote{E-mail: jyo2210@yahoo.co.in}} 

\affiliation{Department Of Physics, Lucknow University, $Lucknow-226007^{\ast,\dagger}$}

\begin{abstract} 
 The proposed Iron Calorimeter (ICAL) detector at Indian Neutrino Observatory(INO) will be a large (50 kt) magnetized detector located 1270 m 
underground at Bodi West Hills in Tamilnadu. ICAL is capable of identifying the charge of the particles. In this paper its potential for 
the measurement of the muon charge ratio is explored by means of a detailed simulation-based study, first using the CORSIKA code and then 
comparing it with an analytical model (the “pika” model). The estimated muon charge ratio is in agreement with the existing experimental 
observations, its measure can be extended by INO-ICAL up to 10 TeV and up to 60 degrees.
\end{abstract}
\maketitle

\section {Introduction}	
\setlength{\baselineskip}{13pt}

 The India-based Neutrino Observatory(INO) is planned to be built at Theni district of Tamilnadu in southern India. This neutrino observatory 
 will have a 52 kton magnetized Iron CALorimeter (ICAL) detector \cite{Shakeel2015}, which is designed to study the physics of atmospheric neutrino flavor 
 oscillations. The main goal of the ICAL detector is to make precise measurements of neutrino oscillation parameters and to determine the neutrino 
 mass hierarchy. Roughly 1.2 km underground, the INO-ICAL detector will be the biggest magnetized detector to measure the cosmic ray muon flux 
 with a capability to distinguish between $\mu^{+}$ and $\mu^{-}$. 
 
 Cosmic rays below energies of about $10^{17}$ eV are thought to be of Galactic origin, the most probable source being Supernova remnants, that 
 constantly hit the outer surface of the Earth's atmosphere. The magnetic field of the Earth tends to exclude low energy charged particles(E $\approx$ 1 GeV), 
 but high energy particles easily manage to penetrate it. These cosmic rays are primarily composed of high energy protons, alpha particles and 
 heavier nuclei, which interact with the top layer of Earth's atmospheric nuclei and produce pions and kaons. These particles further decay to 
 muons \\
 
  \hspace{3cm}    $\pi$$^{-}$ $\rightarrow$ $\mu$$^{-}$ + $\overline\nu$$_{\mu}$  \hspace{3cm}$\pi$$^{+}$ $\rightarrow$ $\mu$$^{+}$ + $\nu$$_{\mu}$ 
        
 \hspace{3cm}  K$^{-}$ $\rightarrow$ $\mu$$^{-}$ + $\overline\nu$$_{\mu}$       \hspace{3cm}        K$^{+}$ $\rightarrow$ $\mu$$^{+}$ + $\nu$$_{\mu}$  \\
 
 The charge ratio of these energetic muons is an important measurable which carries information about (i) $\pi$/K hadronic production ratio 
 (ii) the composition of cosmic ray primaries (iii) the contribution of charmed hadrons. (iv) the neutrino flux at very high energies etc. The 
 measurement of the charge ratio of muons ($\mu+/\mu-$), up to tens of TeV, has been made by several experiments (L3+C\cite{2}, MINOS\cite{3}, 
 CMS\cite{4},  OPERA\cite{5}, etc.) and they have found that charge ratio increases with an increase in energy. As the energy increases, the 
 contribution of muons from kaon decay also increases because the longer-lived pions are more likely to interact before decaying than the 
 shorter-lived kaons, competing processes could be the production of heavy flavors or a heavier primary composition.
        
In this work we present the response of ICAL detector for high energy muon measurements using the latest version of the INO-ICAL code, which is 
based on the Geant4 toolkit\cite{8}. For simulating the charge ratio at the underground detector using the surface muon data, we have added the 
real topography of the hill to the INO-ICAL code. The hill topography is defined using the standard rock chemical composition in the Geant4 code. 
After adding the hill topography in the code energy threshold for INO is estimated by propagating muons vertically from the various depths in the 
rock to the detector. After setting the threshold, muon charge ratio is calculated using cosmic muon flux, which is first generated using CORSIKA(COsmic Rays 
SImulation for KAscade)\cite{7} and then this analysis is repeated with the analytical model known as "pika model"\cite{11}. In CORSIKA, for generating the muon flux only vertical 
primary protons spectrum is considered for our analysis. We have generated all the secondary particles fluxes at the top of the hill and then selected only those muons, 
that have energy above the threshold energy for our analysis. The generated muon flux is used by the Geant4 code as an input flux for muon charge ratio analysis 
by propagating it through the hill to the detector. Muons with energy greater than the threshold energy will enter the detector and produce tracks. Track reconstruction is 
performed using the Kalman Filter Technique\cite{17}. In the central region of each module, the typical value of the magnetic field strength is about 1.5 T in the 
y-direction, which is obtained using the MAGNET6.26 software\cite{6a}. Finally, we estimate the energy and the number of $\mu^{+}$ and $\mu^{-}$ particles which pass 
through rock and reach the detector. For this analysis the ICAL detector momentum and charge resolution are considered. Our estimated results for muon charge ratio 
using CORSIKA and "pika" model are compared with each other as well as with the existing experimental data.

This paper is organized as follows : In section 2, we discuss the ICAL detector geometry, the hill topography and the threshold energy of muons. In section 3, we 
explain the response of high energy muons at INO-ICAL detector and have incorporated the detector efficiency to reconstruct the muon momentum and charge. Section 4 briefly 
explains the cosmic muon flux generation at the surface and the estimation of the muon charge ratio at the underground detector using CORSIKA. This section also 
explains the estimation of the charge ratio using the "pika" model. In section 5 the systematic uncertainities addressed in this work are explained. In section 6 and 7, 
we discuss our results and conclusions respectively. 

\section{THE ICAL detector geometry and the Bodi hill Profile}

\subsection{Detector geometry}
 The INO-ICAL detector will have a modular structure. The full detector consists of three modules each of size 16 m $\times$ 16 m $\times$ 14.45 m. 
 These modules are placed along the x-direction as shown in figure 1, with a gap of 20 cm between them and the origin is taken to be the center of the 
 central module, while the remaining horizontal and transverse direction are labelled as y-axis and z-axis respectively. The z-axis points 
 vertically upwards hence the polar angle is equal to the zenith angle $\theta$. Each module comprises 151 horizontal layers of iron plates of 
 area 16.0 m $\times$ 16.0 m and thickness 5.6 cm with a vertical gap of 4.0 cm, interleaved with Resistive Plate Chamber(RPCs) as an active detector 
 elements. 
 
  \begin{figure}[h!]
\centering
\includegraphics[scale=.4]{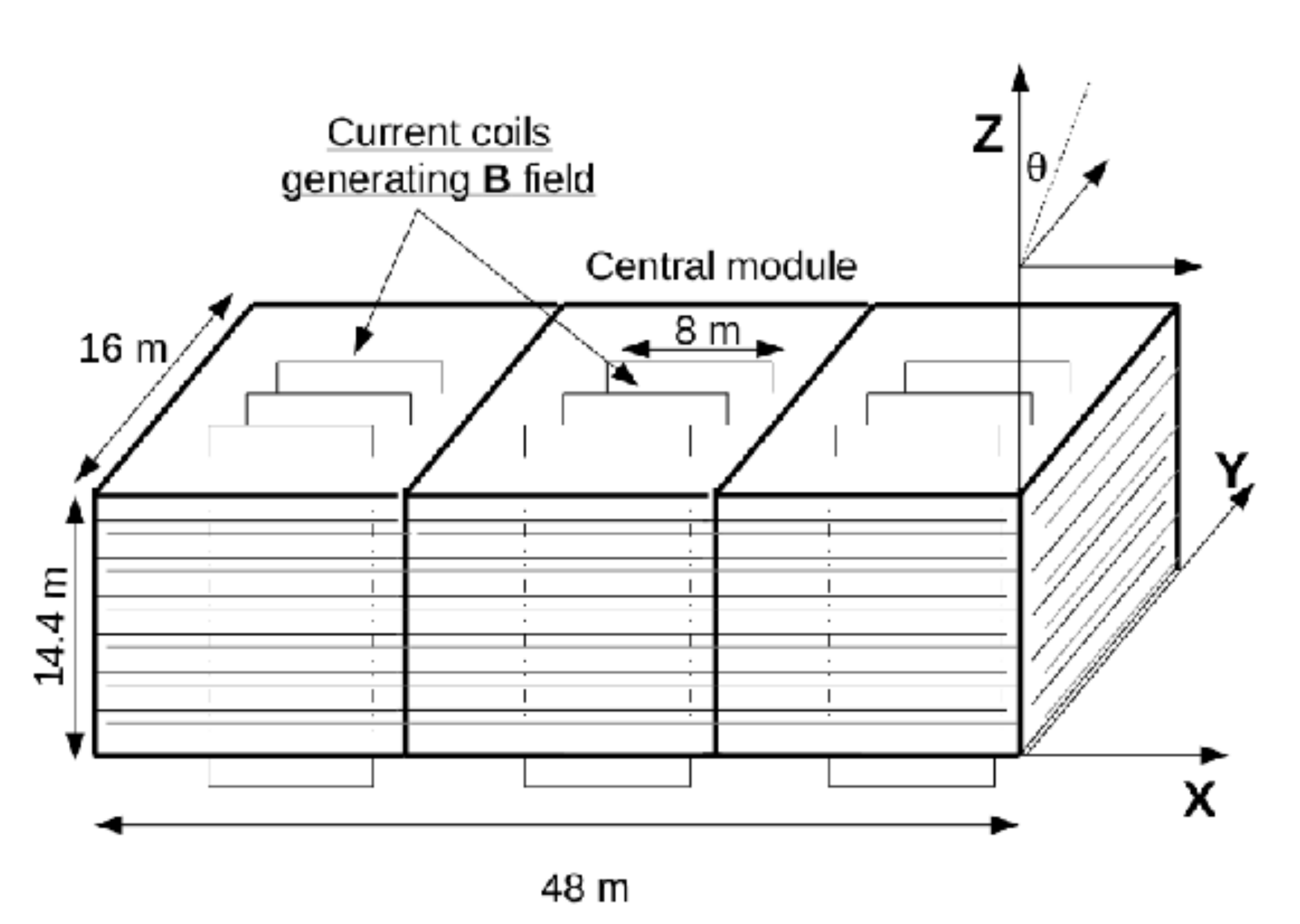}
\caption{\label{fig:ICAL_fig}Schematic view of the INO-ICAL detector geometry.}
\end{figure}
 
 The total mass of the detector is about 52 kton, excluding the weight of the coils. The basic RPC units of size 1.84 m $\times$ 1.84 m $\times$ 
 2.5 cm are placed in grid-format within the air gap \cite{Shakeel2015}. More details of the detector elements and its parameters are discussed in 
 \cite{Shakeel2015}. This geometry has been simulated by the INO collaboration using GEANT4 \cite{8} and its response for muon with energy of few tens of GeV is 
 studied in \cite{9}.

\subsection{Digital data of the INO peak at Pottipuram site.}

 We have generated the elevation profile of the mountain around INO cavern as shown in Fig. 2, at Bodi West Hills in Theni district at Tamilnadu.
\newline The latitude and longitude at the cavern location are as follows:
\begin{itemize}
 \item Latitude : $77^{\circ}15^{'}0^{''}E-77^{\circ}30^{'}0^{''}E$
 \item Longitude : $9^{\circ}57^{'}30^{''}N-10^{\circ}0^{'}0^{''}N$
\end{itemize}
  The average height of the peak around the tunnel region is 1587.32 m from the sea level and the height of the detector location from the sea 
  level is around 317 m, hence the rock cover above the detector is around 1270 m.

  \begin{figure}[h!]
\centering
\includegraphics[scale=0.5]{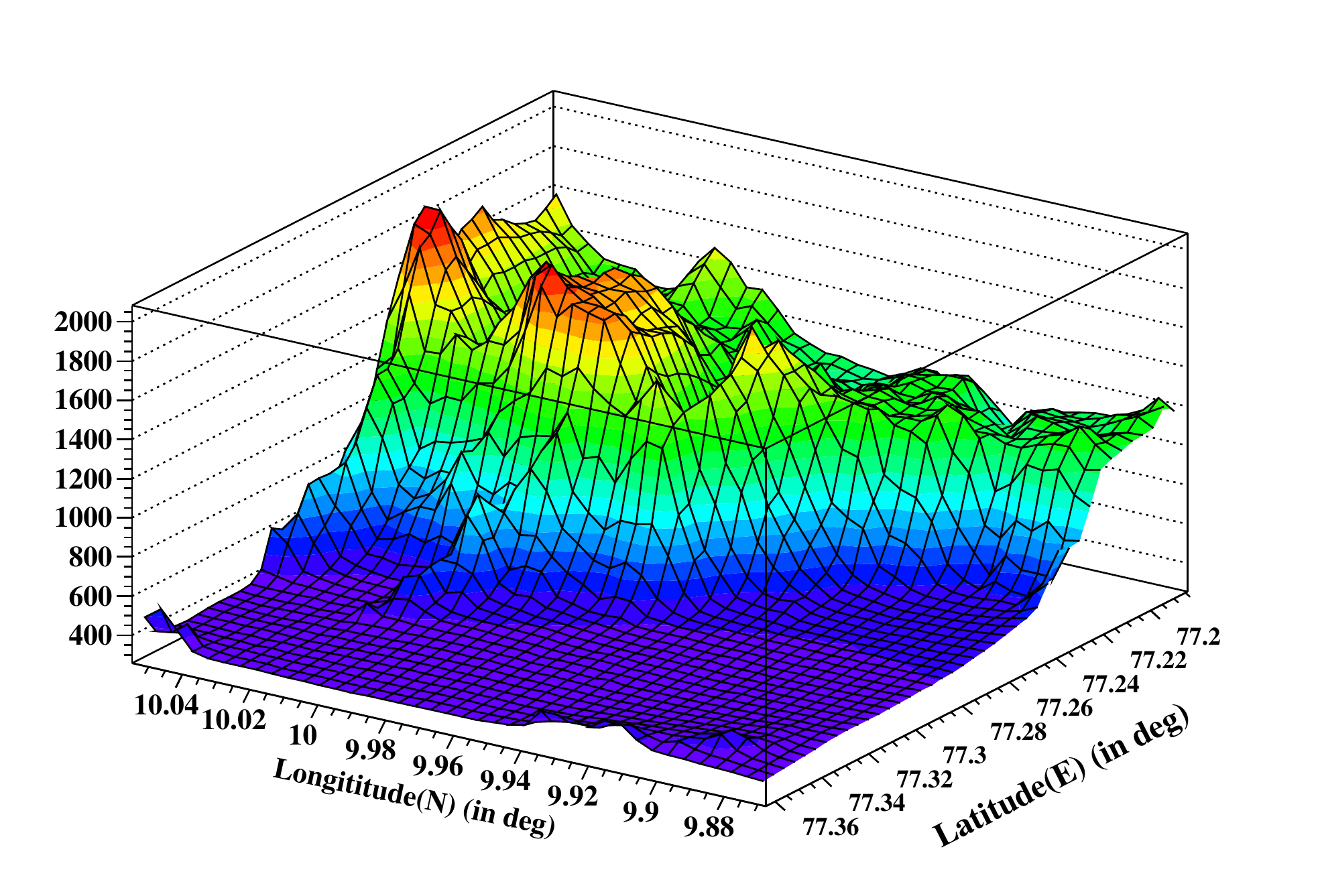}
\caption{\label{fig:rockpro.pdf} 3D plot of mountain surrounding (range 10 km ) around the INO-ICAL cavern location.}
\end{figure}

We are using the mountain profile provided by Geological Survey of India, as shown in Figure 2. For the detailed analysis of underground muon 
charge ratio at INO-ICAL detector, standard rock overburden of density 2.65 gm/cm$^{3}$ is used in our work. The rock type around the INO location is charnockite 
whose density is around 2.6 gm/cm$^{3}$. The real rock density is roughly equal to the standard rock density. 
\subsection{Muon Energy Loss Analysis through Standard Rock:}

The muon charge ratio $R_{\mu^{+}/\mu^{-}}$, measured in underground experiments is the ratio of $\mu^{+}$ to $\mu^{-}$ at the detector. Under the 
hypothesis of identical energy losses for positive and negative muons, the muon charge ratio on the surface would not be modified by propagation of muons
in the rock down to the detector. At higher energies, the fluctuations in energy losses by $\mu^{+}$ and $\mu^{-}$ are important and an accurate calculation of muon 
charge ratio requires a simulation that accounts for stochastic loss processes \cite{23}. Muons of energy $E_{\mu,0}$ at the Earth's surface, lose energy \cite{16} as 
they traverse a slant depth $X$ ($g/cm^{2}$) through the Standard rock to reach the detector. The muon energy loss is described by the well-know Bethe-Bloch formula : 
 \begin{equation}
    -\dfrac{dE}{dX} = a(E_{\mu}) + b(E_{\mu})E_{\mu}
 \end{equation}             
where $a$ parameterizes  the energy loss by collisions and $b$ parameterizes the energy loss by three radiative processes, namely 
Bremsstrahlung, Pair production and Photo-nuclear interactions. The energy loss parameters $a$ and $b$ for the standard rock, as a function of 
energy are taken from reference \cite{16}. The muon energy reconstructed at the detector can be related to the surface muon energy $E_{\mu,0}$ by 
\cite{16}  
  
  \begin{equation}
 E_{\mu} = (E_{\mu,0}+ a/b)e^{bX} - a/b. 
 \end{equation}                
The values of the energy loss parameters $a$ and $b$ depend on the average composition of the rock. Energy loss processes of $\mu^{+}$ and 
$\mu^{-}$ are almost same but there are corrections of the order of the fine structure constant $\alpha$, which were shown to depend almost only 
on the ionisation energy loss \cite{11}. The change $\delta(r_{\mu})$ in charge asymmetry can be written as: 
\begin{equation}
\delta(r_{\mu}) \sim 3.7\delta E_{\mu}/E_{\mu} 
\end{equation}
where $\delta E_{\mu}$ is the difference in energy loss for $\mu^{+}$ and $\mu^{-}$. As shown in \cite{11}, a change from standard rock composition
to a different rock composition produces a negligible effect on the muon charge ratio. As such, our use of standard rock in place of charnockite is justified.
A good knowledge of the rock overburden is required to estimate the muon momentum at the underground detector from the surface muon momentum. The 
variation in the composition of rock requires the muon energy loss estimation in all differential segments of rock, where the density is constant, 
and the addition of these losses give the complete muon energy loss. After feeding the mountain geometry within the ICAL code, 10000 muons were 
propagated vertically to the detector through different depths of the rock shielding. For energy loss analysis in the rock overburden the different 
depths (from the origin of the central module) considered in our analysis are : 1, 2, 3, 3.37, 4, 5 and 6 km.w.e., the value 3.37 km.w.e. refers to 
the highest peak value or maximum rock overburden from the origin of the detector. The muon energy thresholds for increasing slant depths are shown 
in Fig.3, a threshold of 1.6 TeV correspond to the experimental depth of 3.37km.w.e. 
\begin{figure}[htbp]
\centering
\includegraphics[scale=0.5]{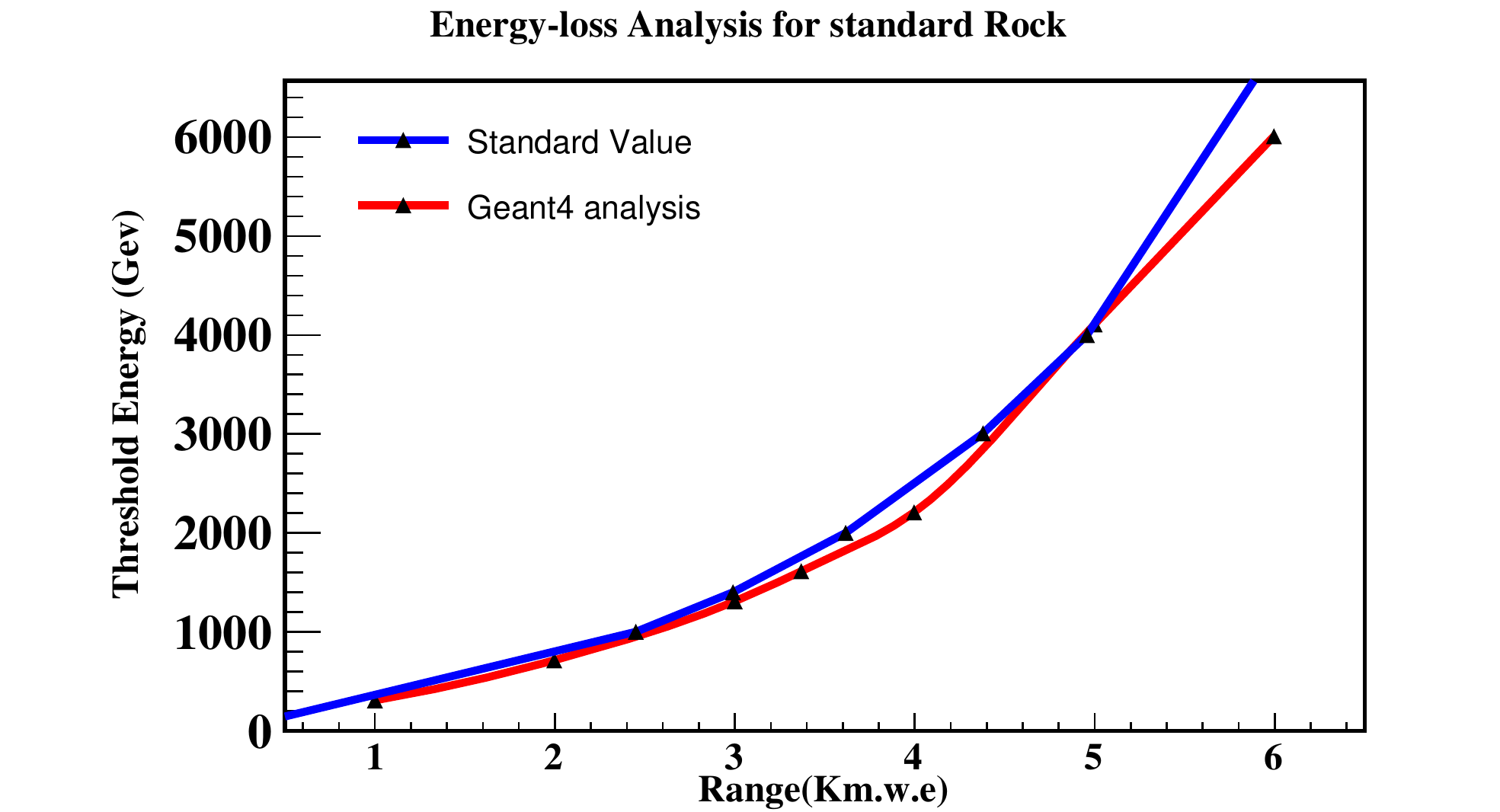}
\caption{\label{fig:enelossanaly}Muon energy threshold versus range in standard rock (blue line). The red line (obtained using the energy loss 
parameter from \cite{16}) is shown for comparison.}
\end{figure}

\section{High Energy Muon Response at ICAL}
In this work we have done simulation for low energy (1-20GeV) muons as well as for the high energy (20-500GeV) muons using the latest version of INO-ICAL code. 
Details of the detector simulation for low energy muons with older version of INO-ICAL code are already published in the reference \cite{9}. In this work we have 
followed the same approach as in reference \cite{9}, for muon response analysis up-to energies 500 GeV inside the detector. Detector response for the peripheral region 
of the detector was performed in \cite{31}, while in this work detector response for the central region is performed. For simulating the response of high energy muon in 
the ICAL detector, 10000 muons were propagated uniformly from a vertex, randomly located inside the central volume of the detector of dimensions 8 m $\times$ 8 m $\times$ 10 m. 
In our simulation we have considered a uniform magnetic field of strength 1.5T. 

\subsection{Tracks and Event selection}

The muon track reconstruction is based on a Kalman filter algorithm that takes into account the local magnetic field. This algorithm is briefly described in \cite{9}. 
It is used to fit the tracks  based on the bending of the tracks in the magnetic field, by collecting the hitting points in multiple layers of iron plates. Only tracks 
for which the quality of fit is better than $\chi^{2}/ndf <$ 10 are used in this analysis. Each muon track is further analysed in order to identify its direction, 
charge and momentum. Both, fully contained and partially contained muons events are included in this work.

\subsection{Momentum Reconstruction Efficiency of ICAL}

The momentum reconstruction efficiency($\epsilon_{rec}$) is defined as the ratio of the number of reconstructed events, n$_{rec}$, to the total 
number of generated events, N$_{total}$. We have \\
\begin{equation}
\epsilon_{rec} = \boldmath\frac{{n_{rec}}}{\boldmath N_{total}}
\end{equation}                 
  with uncertainity, 
  \begin{equation}
\delta \epsilon_{rec}=\boldmath\sqrt{(\epsilon_{rec}(1-\epsilon_{rec})/\boldmath N_{total})}
 \end{equation}
Figure 4 shows the muon momentum reconstruction efficiency as a function of input momentum for different cos$\theta$ values. 
\begin{figure}[htbp]
\centering
\includegraphics[scale=0.7]{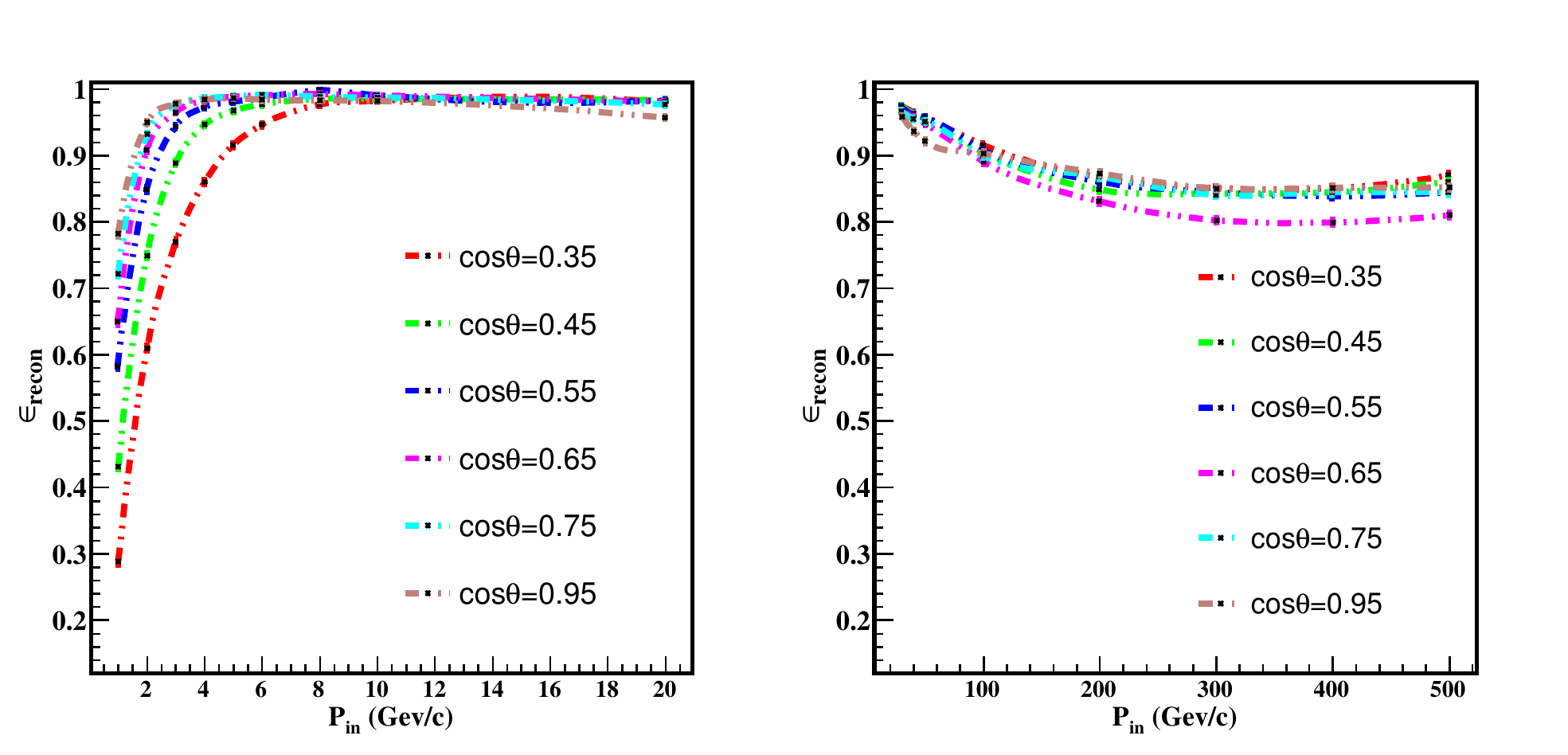}
\caption{\label{fig:recmombotherror}Muon momentum reconstruction efficiency as a function of the input momentum for different zenith angles. The left 
panel refers to low energy, 1$\leq$ P$_{in}$ $\leq$20GeV muons, while the right panel refers to the high energy, 20$\leq$ P$_{in}$ $\leq$500 GeV 
muons.}
\end{figure}
The left and right panels demonstrate detector response for low and high energy muon momentum respectively. One can see that the momentum reconstruction efficiency 
depends on the incident particle momentum and the angle of propagation. The rise in efficiency of muons , as shown in the left panel of figure 4, is due to the fact 
that higher energy muons can cross more number layers. A muon with fixed energy hitting the detector at larger zenith angle will give lesser no of hits as compared to 
the same muon hitting the detector vertically (angle dependence). At higher energies the reconstruction efficiency of muons, as shown in the right panel of figure 4, initially drops 
and then becomes almost constant (around 80$\%$). The reason for the drop in momentum reconstruction efficiency of muons at high energies (above about 50 GeV) is, that 
the muons travel nearly straight without being deflected in the magnetic field of the detector. 

Charge ratio of muons typically depends on the reconstructed muon momentum at the detector, hence a change in momentum value will lead to an uncertainity 
in the charge ratio. Figure 5 indicates a shift in the mean reconstructed muon momentum distribution at fixed zentith angle and this shift increases with 
increasing input momentum. Figure 6 shows the shift in muon momentum (P$_{in}$-P$_{rec}$) as function of input muon momentum for various zenith angles. This shift can 
arise due to multiple scattering and it increases faster in the higher energy range.  

 \begin{figure}
\centering
\includegraphics[scale=0.6]{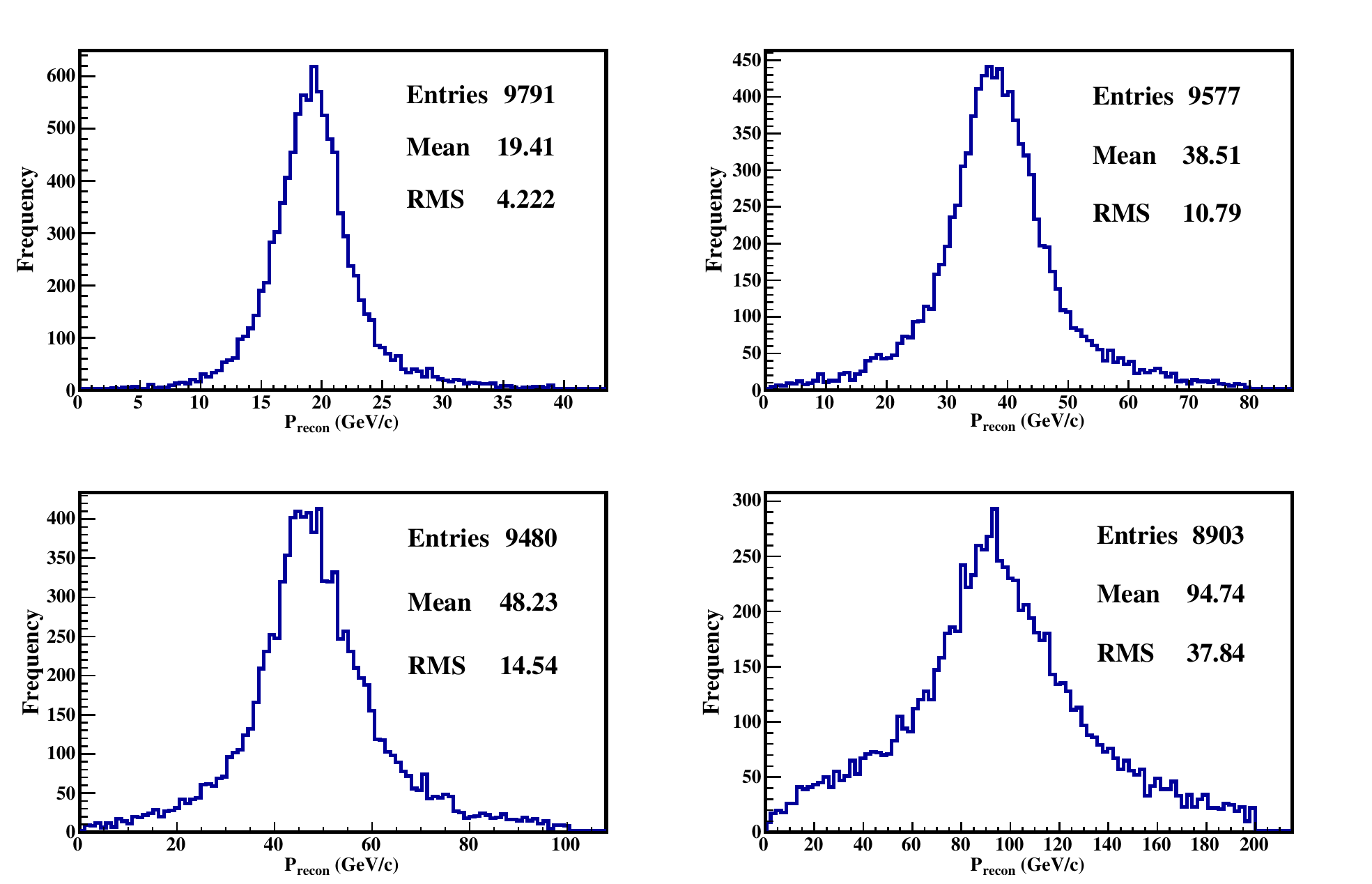}
\caption{\label{fig:shiftmom}Reconstructed muon momentum distribution for four different values(20, 40, 50 and 100 GeV) of the input muon momentum and at fix zenith angle,
$cos\theta$=0.65}
\end{figure}

\begin{figure}
\centering
\includegraphics[scale=0.5]{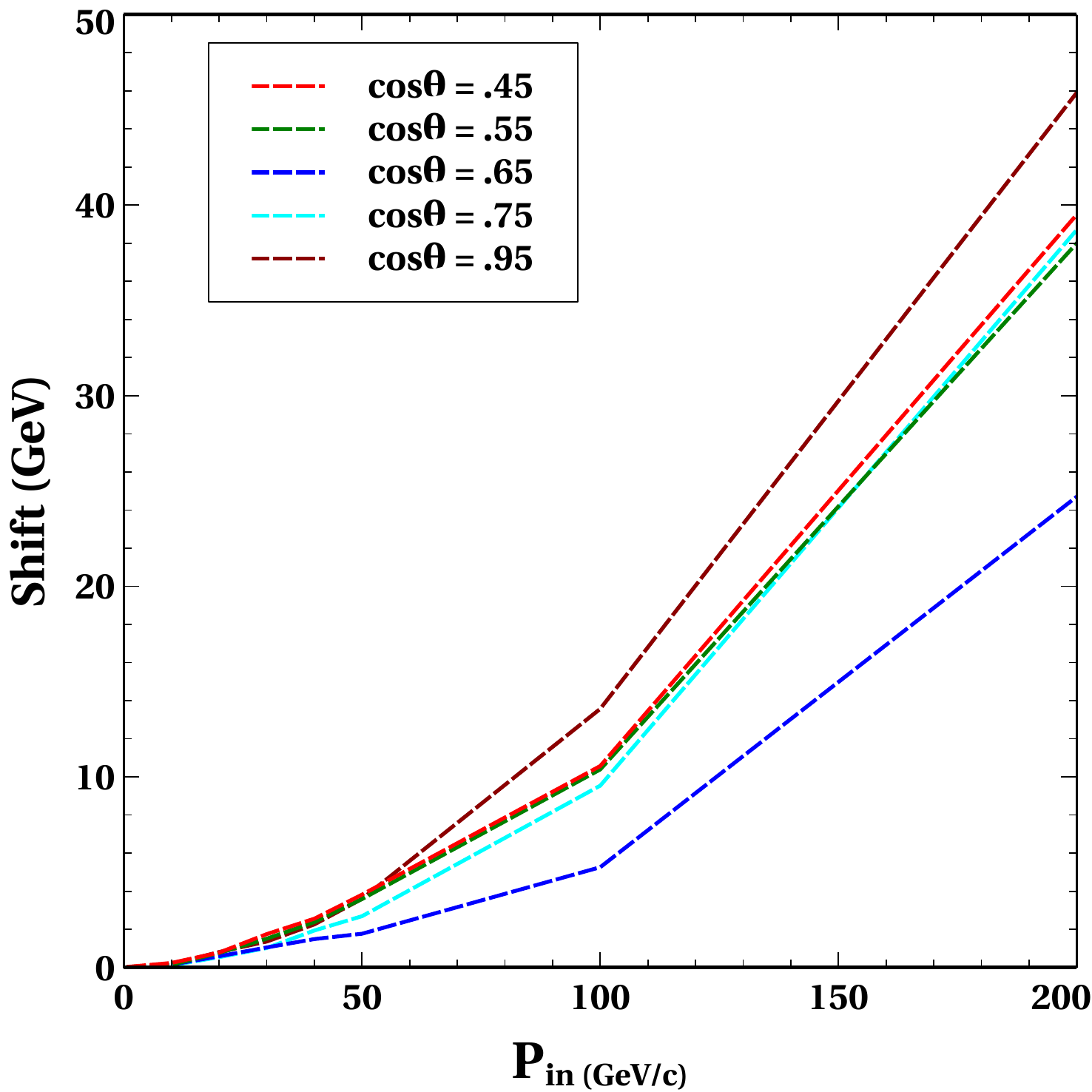}
\caption{\label{fig:momshiftplot}Shift in the mean of the reconstructed momentum as a function of the input momentum.}
\end{figure}

\subsection{Relative Charge Identification Efficiency of ICAL}

\begin{figure}[htbp]
\centering
\includegraphics[scale=0.7]{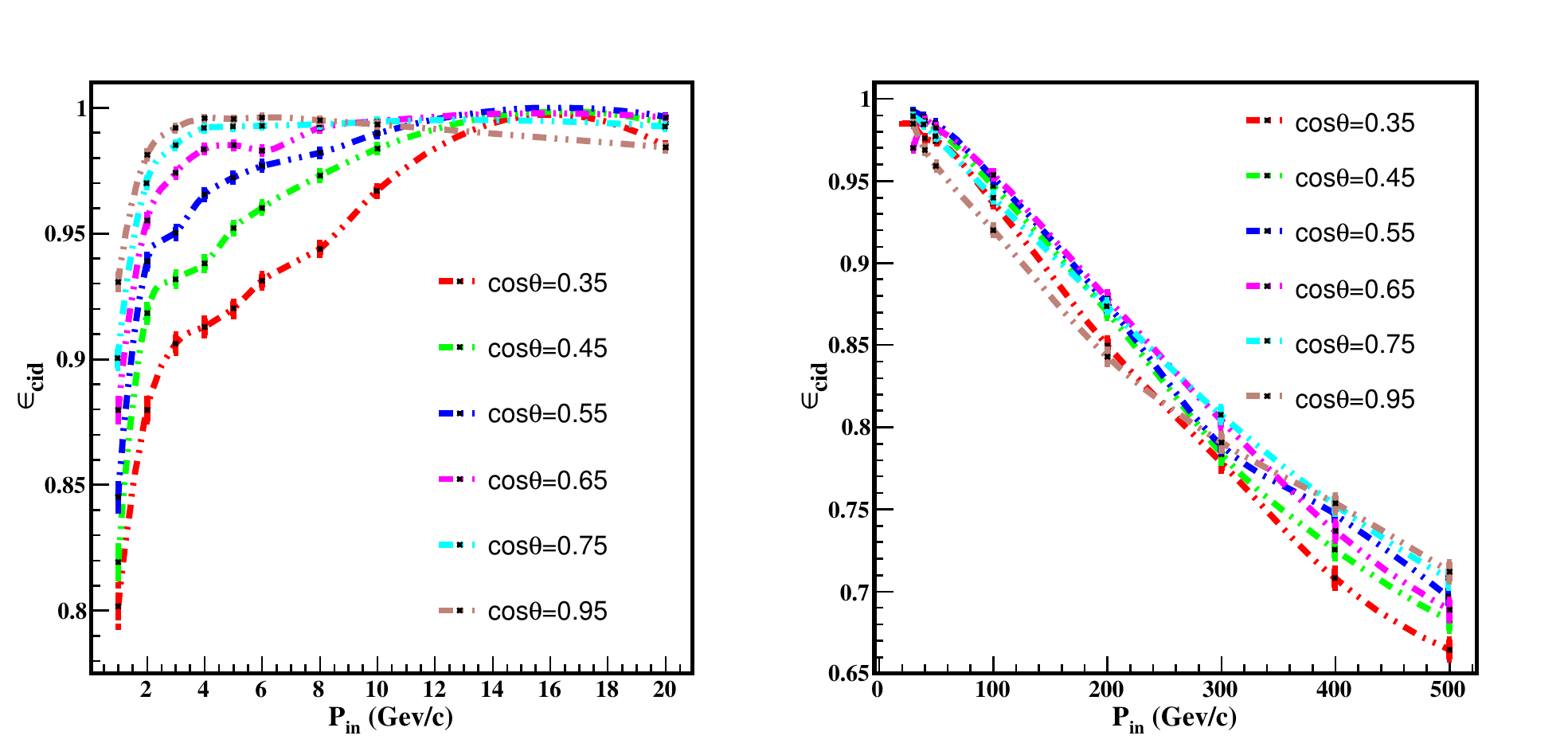}
\caption{\label{fig:reccidbotherr}The relative charge identification efficiency as a function of the input muon momentum for different cos$\theta$ 
values. The left panel refers to low energy, 1$\leq$ P$_{in}$ $\leq$20 GeV muons while the right panel refers to the high energy, 20$\leq$ P$_{in}$ $\leq$500 GeV muons.}
\end{figure}

The charge of the particle is determined from the direction of curvature of the particle track in the magnetic field. Thus it is crucial for 
the determination of the neutrino mass hierarchy, which is the main aim for the INO project, and for the estimation of the atmospheric muon 
charge ratio. Relative charge identification efficiency is defined as the ratio of the number of events with correct charge identification, 
n$_{cid}$ to the total number of reconstructed events, n$_{rec}$ of same charge i.e.,
 \begin{equation}
 \epsilon_{cid} = \boldmath \frac{n_{cid}}{\boldmath n_{rec}}
 \end{equation}       
 with uncertainity,   
 \begin{equation}
 \boldmath\delta \epsilon_{cid}\boldmath=\boldmath\sqrt{(\epsilon_{rec}(1-\epsilon_{cid})/\boldmath n_{rec})}. 
  \end{equation}

Figure 7 shows the relative charge identification efficiency as a function of input momentum for different cos$\theta$ values. Here the left and 
right panels demonstrate the detector response for low and high energy muon momentum respectively. Muons propagated with small momentum will cross lesser number of 
layers and will give lesser number of hits. This may lead to an incorrect reconstruction of the direction of bending and wrong 
charge identification. Hence the charge identification efficiency is relatively poor at lower energies, as can be seen from the 
left plot of Figure 7. As the energy increases, the length of the track also increases due to which the charge identification efficiency improves. 
Beyond a few GeV/c, the charge identification efficiency becomes roughly constant at 98-99$\%$. But as the energy increases further above 50 GeV 
the charge identification efficiency starts decreasing because in the constant magnetic field the deflection of highly energetic muon will be less, 
and at a certain energy the muon will leave the detector without getting deflected by magnetic field. Due to this limitation the charge 
identification efficiency falls to 70$\%$ at muon momentum 400 GeV/c. In our work we have simulated the muon charge ratio using CORSIKA upto muon 
momenta of 400 GeV/c, where the detector charge identification efficiency is $\geq$ 70$\%$ for the central region. The 70$\%$ efficiency of the detector is only 
for the muons generated by vertical protons using CORSIKA, where as the same studies have been extended in section 4.2 for 90$\%$ efficiency. 

\section{The muon charge ratio evaluation:}
\subsection{Muon Charge Ratio Analysis using CORSIKA data:}
CORSIKA, is used here for air shower simulation. We exploited QGSJET II-04 \cite{07} for high-energy 
interactions and GHEISHA\cite{7a} for low energy. Preliminary tests with a different low energy interaction code showed negligible difference 
in the final determination of the muon charge ratio.

We have generated the muon flux using CORSIKA at an observational level of 2.0 Km. above the sea level, which is the top surface of the hill at the 
INO location as shown in Fig. 2. 
\begin{figure}[htbp]
\centering
\includegraphics[scale=0.6]{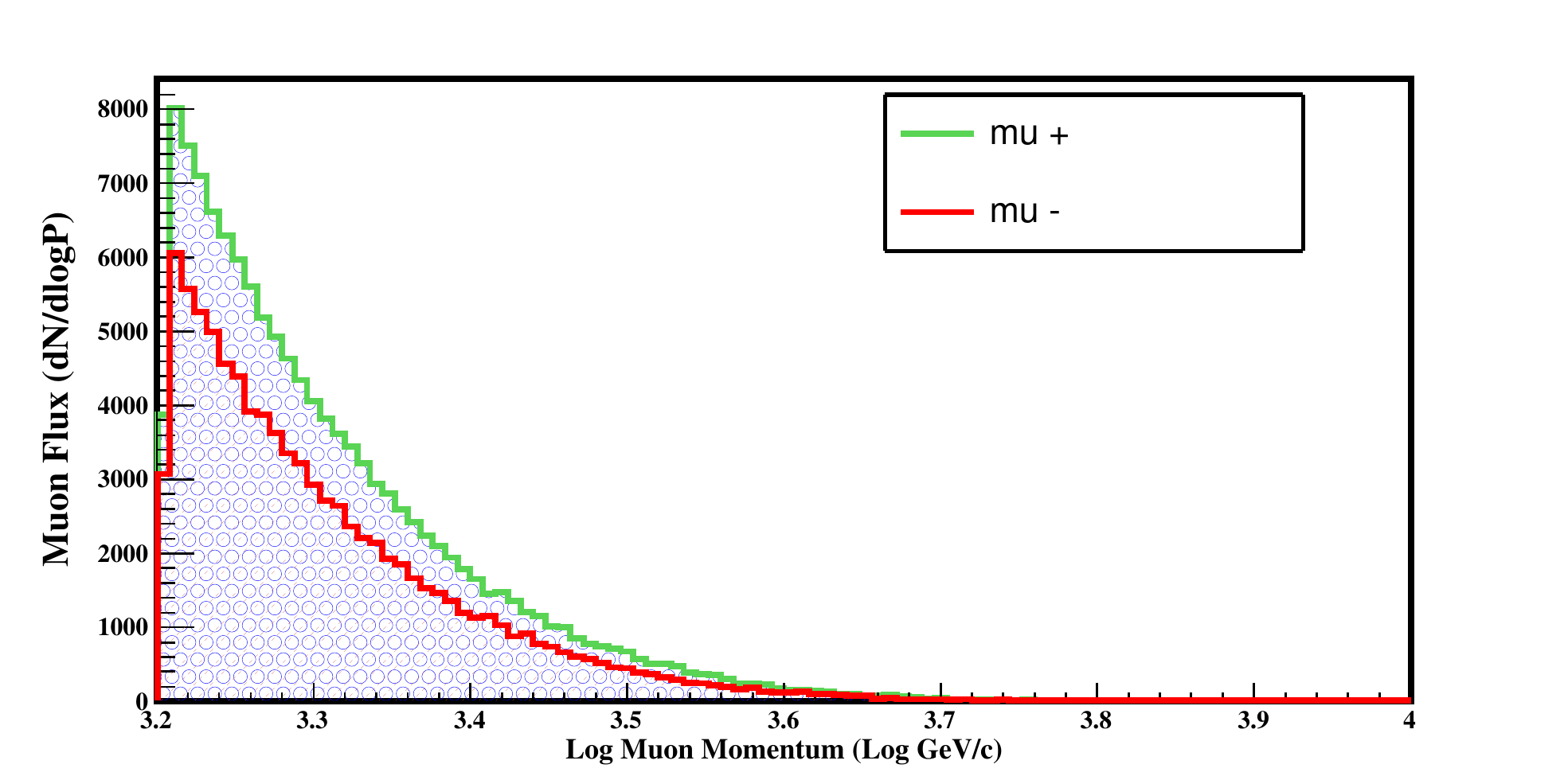}
\caption{\label{fig:verticalmuonsflux.pdf}Differential Muon flux generated using CORSIKA at the top of the hill as a function of muon momentum at log 
scale.}
\end{figure}
\begin{figure}[htbp]
\centering
\includegraphics[scale=0.6]{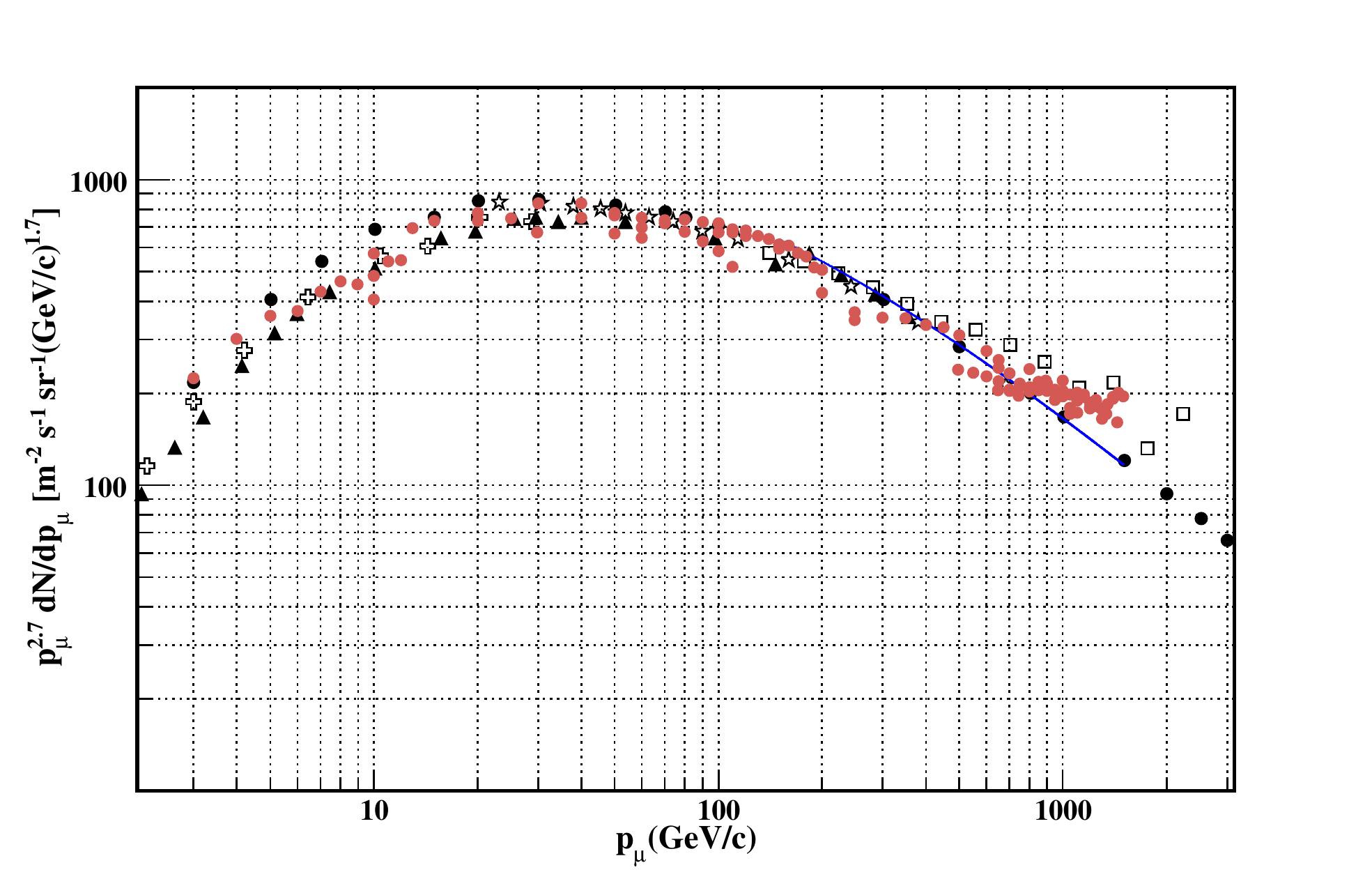}
\caption{\label{fig:ExpCorsikaplot2}Muon flux generated using CORSIKA at the top of the hill (i.e. 2 k.m. above the sea 
level) with primary protons in energy range 1-20 TeV(brown dots) and its comparison with the existing experimental fluxes ($\bigstar$ \cite{2},
$\vartriangle$\cite{6}, +\cite{18}, $\square$\cite{19}, $\bullet$\cite{20}) and the flux (solid blue line) from \cite{25}.}
\end{figure}
This muon flux is used as an input for GEANT4 based INO-ICAL code. Muon energy loss analysis in the rock suggests that only high energy muons will 
be able to reach the detector, so for the cosmic muon analysis only high energy muons are generated for this work. For the generation of 
primary proton flux above the atmosphere, a power law parameterization is considered($E^{-2.7}$). Proton flux is generated at 0$^{\circ}$ 
zenith angle in the energy range 1-20 TeV. The entire energy range is divided into 100 equal bins with the bin size of 200 GeV and each 
bin has 1000000 protons. The generated muon spectrum from the selected primary spectrum is shown in Fig.8 as a function of muon momentum using a 
log scale. The expected time exposure for collecting this data is around 3 years \cite{15}. To validate the surface muon flux data generated by us 
using CORSIKA, we have compared our flux with the muon flux evaluated by T. Gaisser \cite{25} and the available experimental data. This is illustrated in Figure 9 
as a function of muon momentum (only vertical muon flux is used). The generated muon flux at the top of the hill surface will spread in the large area of X-Y plane 
as shown in Figure 10 where the observational plane is at 2.0 km. above the sea level. Most of the muons will be distributed in the range -100 m $\leq$ X $\leq$  100 m 
and -100 m $\leq$ Y $\leq$ 100 m of X-Y plane and the origin of the plane will be the shower axis. 

\begin{figure}[htbp]
\centering
\includegraphics[scale=0.5]{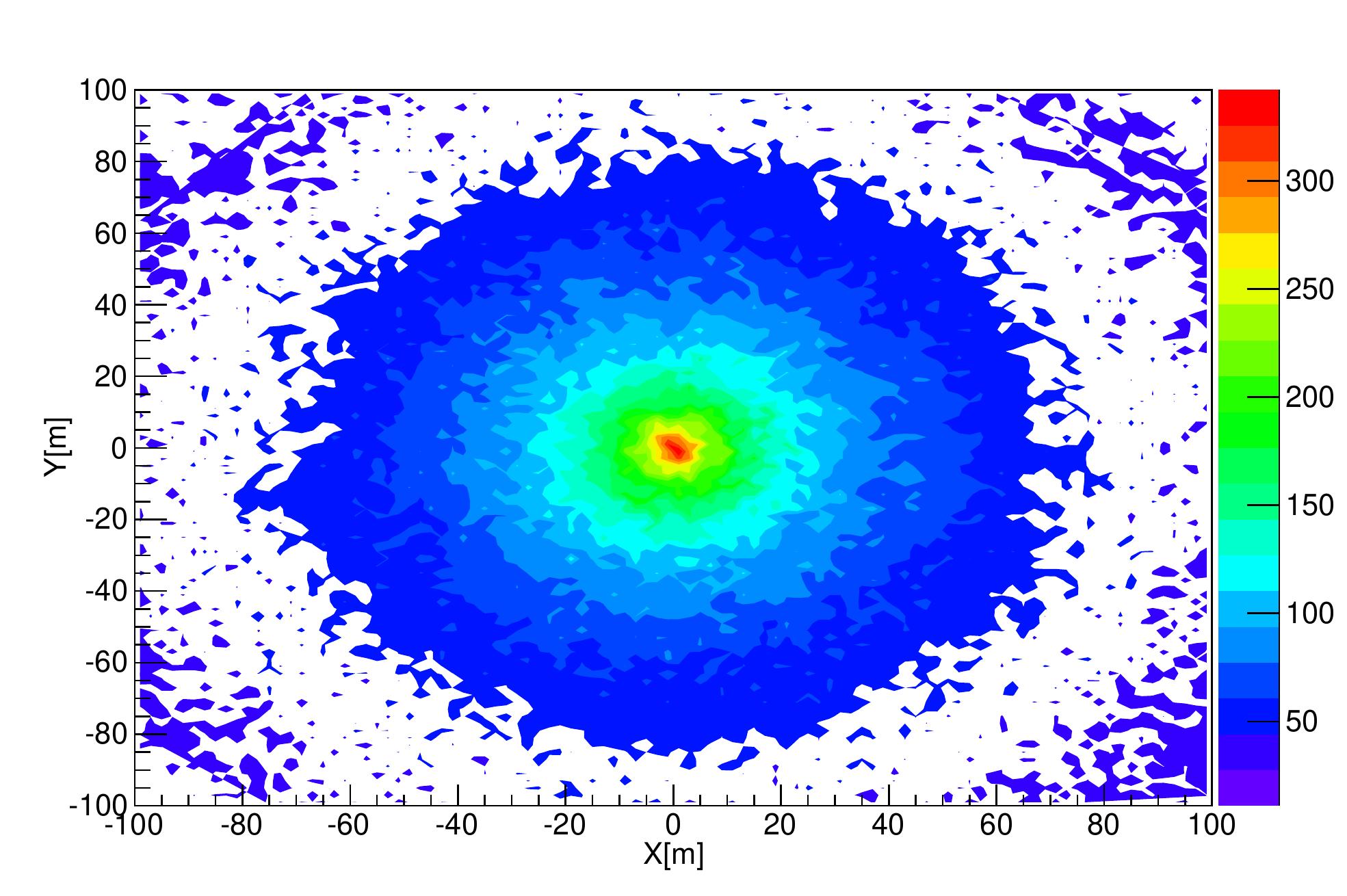}
\caption{\label{fig:pos_plotin_meter} The distribution of secondary muons produced 2.0 Km. above the sea level from the vertical primary proton 
shower, in X-Y plane.}
\end{figure}

Density of the muons near the shower axis is very high as shown in Figure 10 and the uncertainity in the distribution of the flux is considered 
in the statistical error bars of the final results.

In this section we determine the muon charge ratio, $R_{{\mu^{+}}/{\mu^{-}}}$$\equiv$$N_{\mu^{+}}$/$N_{\mu^{-}}$, as a function of the 
reconstructed muon momentum at the INO-ICAL detector. 
  
\begin{figure}[htbp]
\centering
\includegraphics[scale=0.4]{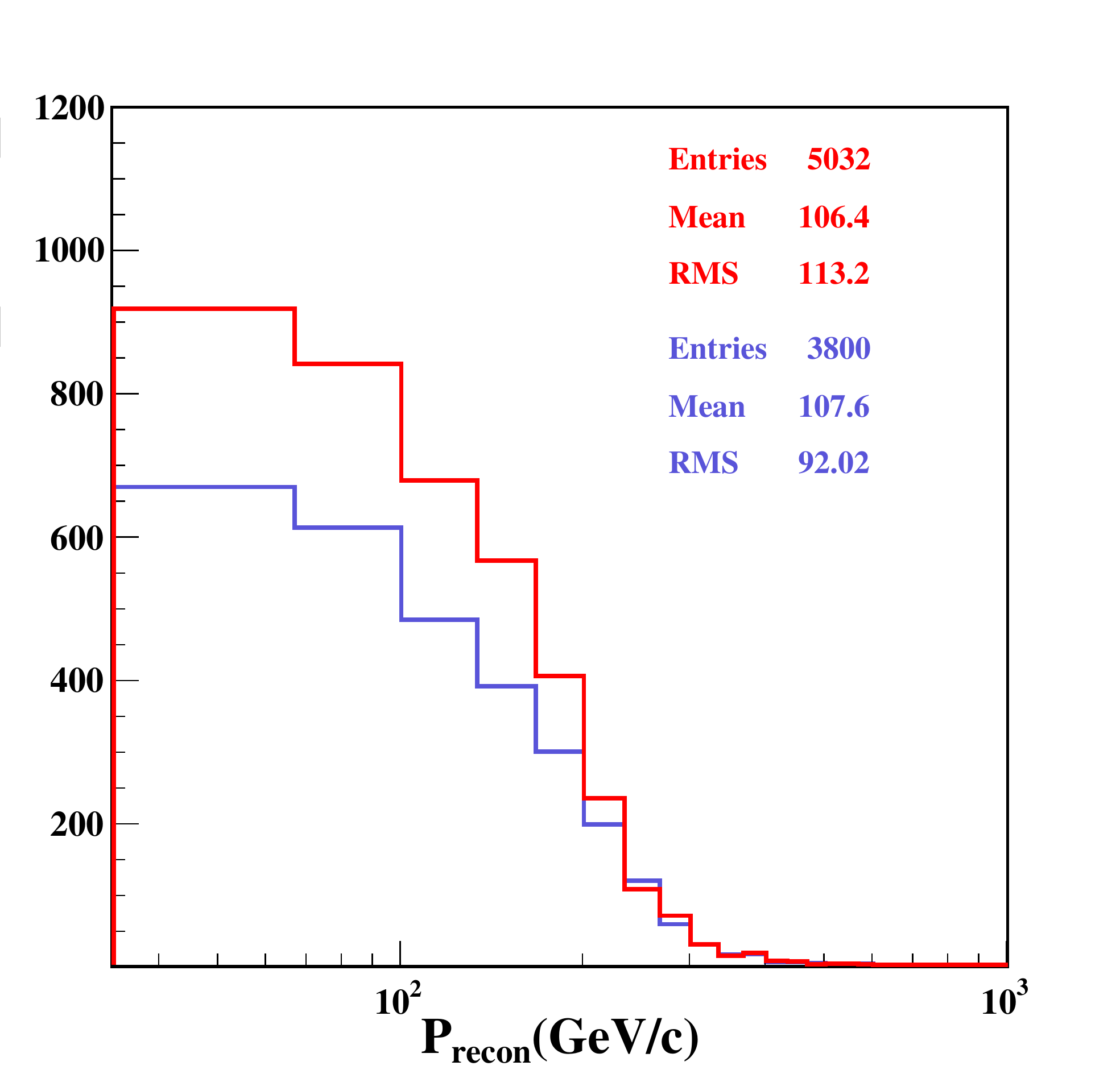}
\caption{\label{fig:recmomdata1}Differential muon flux(red line $\mu^{+}$ and blue line $\mu^{-}$) at the ICAL detector as a function of 
reconstructed muon momentum after passing through the rock.}

\end{figure}

\begin{table}[htp]
\caption{DATA TABLE FOR CHARGE RATIO ANALYSIS FROM VERTICAL MUONS.}
\renewcommand\thetable{\Roman{table}}
\centering
\setlength{\tabcolsep}{2pt}
\begin{tabular}{|c | c | c | c | c | c | c | c |}
\hline
 Energy (TeV)& $R_{\mu+/\mu-}$ at ICAL & $R_{\mu+/\mu-} ^{Surface}CORSIKA$ & $R_{\mu+/\mu-}^{Surface}PIKA $ \\
\hline\hline
  1.60-1.65 &   1.32$\pm$0.013   &  1.31 & 1.41\\
  1.65-1.70 &   1.31 $\pm$0.022  &  1.34 & 1.41\\

 1.70-1.75  &   1.32$\pm$0.014    &  1.34& 1.41\\
 1.75-1.80  &   1.32$\pm$0.034    &  1.37 &1.41\\
 1.80-1.85  &   1.41$\pm$0.008    &  1.40 & 1.41\\
 1.85-1.90  &   1.36$\pm$0.008    &  1.37 & 1.41\\
 1.90-1.95  &   1.41$\pm$0.026    &  1.37 &1.41\\
 1.95-2.00  &   1.32$\pm$0.030    &  1.36 &1.41 \\

\hline
\hline
$\langle{R_{\mu+/\mu-}}\rangle$ & 1.35$\pm$0.019&1.36 & 1.41\\
\hline
\end{tabular}
\end{table}  

  From the generated muon data sample only energetic muons with energies in the range 1600GeV to 2000GeV are selected. This muon flux 
  sample is then divided into 8 equal data samples and each have bin size of 50 GeV. Each data sample is propagated through the rock to the 
  detector. The reconstructed muon momentum distribution for one of the sets is shown in Fig.11. This figure shows a clear separation of 
  $\mu^{+}$ and $\mu^{-}$ flux as a function of reconstructed muon momentum at the detector. The corresponding energy of the muons at the surface 
  is 1600-1650GeV. 
  
  \begin{figure}[htbp]
\centering
\includegraphics[scale=0.5]{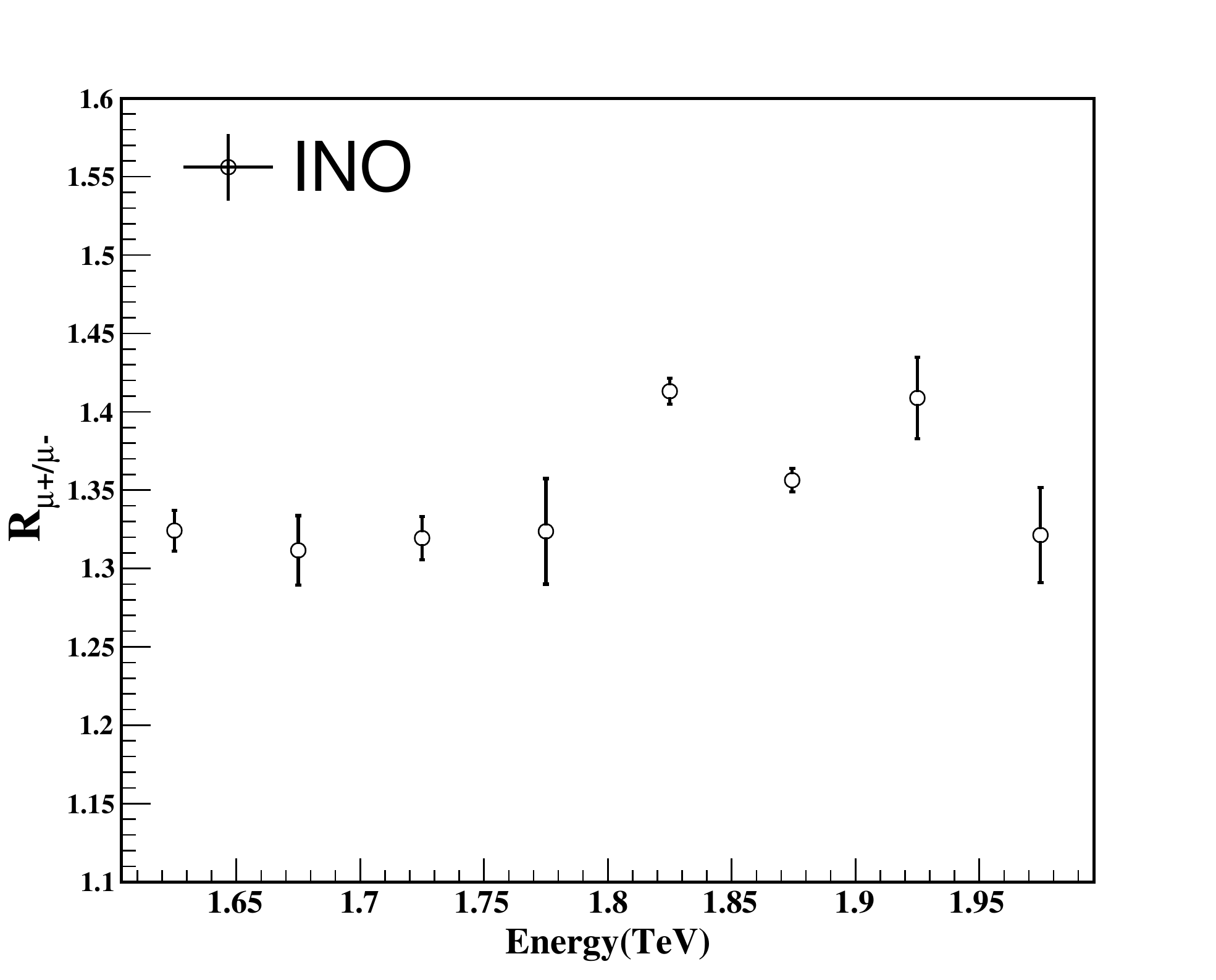}
\caption{\label{fig:Inochargeratio2} Muon charge ratio with statistical errors at the INO-ICAL detector as a function of surface muon energy, obtained using the CORSIKA data.}
 \end{figure} 
 
 The vertical muon charge ratio has been evaluated up to 2 TeV and plotted in Figure 12 and the corresponding values are listed in the second columns 
 of Table 1, whereas the third columns shows the vertical muons charge ratio at the surface. The error bar shown in the Figure 12 represents statistical errors only.


\subsection{Charge ratio analysis using "pika" model:}
 Since the muon charge ratio also depends on the zenith angle, hence we study it as function of $E_{\mu}cos{\theta}$. In this section, the muon 
 charge ratio at INO-ICAL detector is estimated using atmospheric muon energy spectrum \cite{11}. A formula for atmospheric muon energy spectrum 
 given by Gaisser \cite{25} is,
 \begin{equation}
\dfrac{dN_{\mu}}{dE_{\mu}} = \dfrac{0.14E_{\mu} ^{-2.7}}{cm^{2}\hspace{1mm} s\hspace{1mm} sr \hspace{1mm} GeV}\times \bigg \{{\dfrac{1}{1+\dfrac{1.1E_{\mu}cos\theta}{\epsilon_{\pi}}}}+\dfrac{\eta}{1+\dfrac{1.1E_{\mu}cos{\theta}}{\epsilon_{K}}} \bigg \}
 \end{equation}
      
This formula is valid when muon decay is negligible $(E_{\mu} > 100/cos{\theta}$ GeV) and the curvature of the Earth is neglected $(\theta < 70^{0})$. 
The two terms inside the curly braces give the contributions to muon flux from charged pion and kaons. The rise in the muon charge ratio can be 
understood from the properties of $\pi$ and K mesons, and the observation of the rise in muon charge ratio can be used to determine the 
$\pi^{+}/\pi^{-}$ and $K^{+}/K^{-}$ ratio. A method to study separately the positive and negative muons intensities has been developed in 
reference \cite{11} and it is known as "pika" model. This model provides the muon energy spectrum for +ve and -ve muon by using the positive fraction parameters 
$f_{\pi}$ and $f_{K}$,
 
 \begin{equation}
\dfrac{dN_{\mu^{+}}}{dE_{\mu}}= \dfrac{0.14E_{\mu} ^{-2.7}}{cm^{2}\hspace{1mm} s\hspace{1mm} sr \hspace{1mm} GeV}\times \bigg \{{\dfrac{f_{\pi}}{1+\dfrac{1.1E_{\mu}cos\theta}{115 GeV}}} +\dfrac{\eta \times f_{K}}{1+\dfrac{1.1E_{\mu}cos{\theta}}{850 GeV}} \bigg \}  
 \end{equation}
      
 \begin{equation}
\dfrac{dN_{\mu^{-}}}{dE_{\mu}}=\dfrac{0.14E_{\mu} ^{-2.7}}{cm^{2}\hspace{1mm} s\hspace{1mm} sr \hspace{1mm} GeV}\times \bigg \{{\dfrac{1-f_{\pi}}{1+\dfrac{1.1E_{\mu}cos\theta}{115 GeV}}}+\dfrac{\eta \times (1-f_{K})}{1+\dfrac{1.1E_{\mu}cos{\theta}}{850 GeV}} 
  \bigg \}
\end{equation}
      
   In the above equation $\epsilon_\pi$ and $\epsilon_K $ have been replaced by their numerical values. The parameter $\eta$ sets the relative 
   contribution to the muon flux from $\pi$ and K decay. It depends upon the $\pi/K$ ratio, branching ratios and kinematic factors which arise due 
   to difference between the $\pi$ and K masses. The numerical value of this parameter($\eta$ = 0.054) is discussed in \cite{29}. Equations (9) and 
   (10) give the expression for the surface muon charge ratio:

\begin{equation}
 R_{\mu}= \dfrac {\bigg \{ {\dfrac{f_{\pi}}{1+\dfrac{1.1E_{\mu}cos\theta}{115 GeV}}}+ \dfrac{\eta \times f_{K}}{1+\dfrac{1.1E_{\mu}cos{\theta}}{850 GeV}} \bigg \}}
 {\bigg \{ {\dfrac{1-f_{\pi}}{1+\dfrac{1.1E_{\mu}cos\theta}{115 GeV}}} + \dfrac{\eta \times (1-f_{K})}{1+\dfrac{1.1E_{\mu}cos{\theta}}{850 GeV}}\bigg \}}
 \end{equation}
  \begin{figure}
\centering
\includegraphics[scale=0.5]{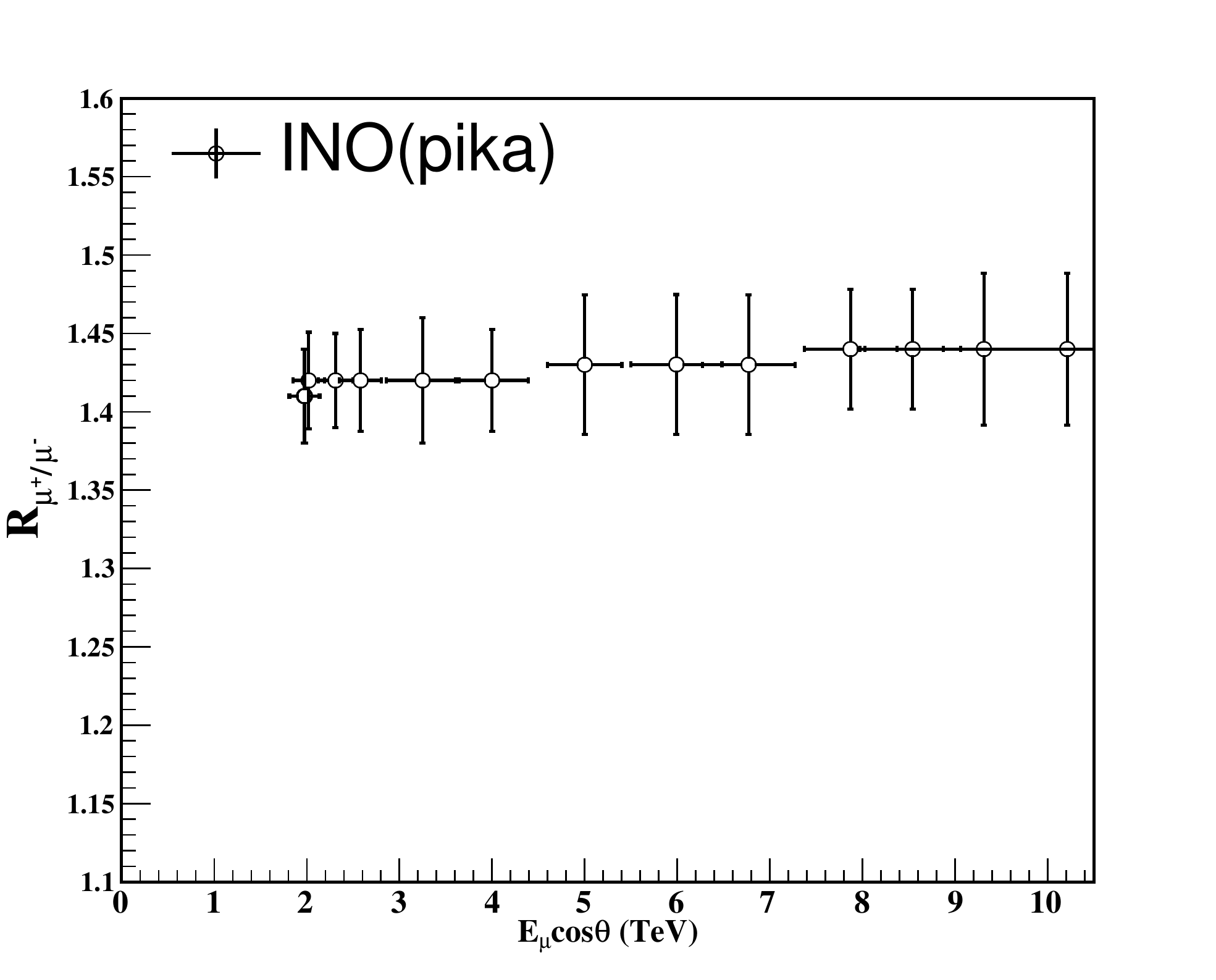}
\caption{\label{fig:munratiocosplot} Muon charge ratio with systematic errors at the INO-ICAL detector as a function of surface muon energy, obtained using the "pika" 
model.}
\end{figure}
       
 The charge ratio of muon from pion decay is $r_{\pi}$ = $f_{\pi}/(1-f_{\pi})$ and from kaon decay is $r_{K} = f_{K}/(1-f_{K})$. Equation (11) is 
 a function of $E_{\mu}cos\theta$ only, all other parameters are extrapolated from the available experimental data. The pika  model describes  
 the data well \cite{11}, from the experiment L3+C \cite{2}, MINOS Near and Far Detector\cite{3}, in comparison to the calculations made by 
 Honda \cite{26}, CORT \cite{27} and Lipari \cite{28}. The charge ratio estimated at the INO detector is shown in Figure 13, for charge ratio 
 calculation with the pika model muons upto momentum of 150 GeV/c at the detector are considered where the charge id efficiency is roughly 90$\%$. The fourth 
 column of Table-1 shows the charge ratio for vertical muons only. Muon flux generation using the Gaisser formula at INO and muon charge ratio estimation is 
 performed in \cite{30}.

\section{Systematic uncertainity:}
    The simulation technique used for the estimation of the muon charge ratio tends to cancel out many systematic uncertainties, however few 
    uncertainties which do not cancel out remain in our analysis. The systematic uncertainties considered in our work are: (I) those 
    arising due to the range estimation in the standard rock for $\mu^{+}$ and $\mu^{-}$ as discussed in section 2.3. (II) those arising due to 
    the shift in the mean of the reconstructed muon momentum as discussed in section 3.3. (III) those arising due to the track reconstruction 
    algorithms in magnetic field, which comes mainly due to the non-uniformity in the magnetic field and available dead spaces as discussed in 
    detail in reference \cite{9}. (IV) those arising due to variation in the surface height,(as stated, we did our analysis for a fixed height 
    (1300m) and then varied this height by $\pm$ 50 m to get the error in charge ratio results due to this variation). The total average systematic 
    uncertainties as discussed above are 0.055, which is included in our error bars as shown in fig. 13.
    
    To evaluate the systematic uncertainities listed above, we exploited a monte carlo code for the estimaton of the slant depth, in unit 
    of kilometer water equivalent (where 1 km.w.e = $10^{5}$ g/cm$^{2}$). For the estimation of the accurate slant depth, we have zoomed 3.0 km 
    of the surrounding area about the origin of the detector location. We observe that the height of the rock cover in this region, is around $1300\pm50$ m above 
    the detector. Equation (2) shows a relationship between slant depth ($X_{I}$) and threshold energy of muon to reach the detector after travelling a distance $X_{I}$. 
    The values of parameters $a$ and $b$ in equation (2) depends on muon energy and the values of these parameters at different muon energies are 
    taken from \cite{16}. Using these energy dependent values of $a$ and $b$ we have estimated the threshold (for muon) for the corresponding slant 
    depth covering in the zenith angle range $0^{0}-60^{0}$. In our simulation we consider only downgoing muons and assume constant height ($1300\pm50$ m). 
    The azimuthal angle is varied to cover all the 5 planes which include the top and four sides of the detector, excluding the bottom surface. This simulation 
    is limited to down going muons therefore the bottom surface of the detector is excluded. Detector efficiency is more than 80$\%$ in the energy range of 1-200 GeV 
    for the zenith angle $(0^{0}-70^{0})$. In our work charge ratio estimation shown in Figure 13 and 14 corresponds to the zenith angle range $0^{0}-60^{0}$. The 
    surface near the region of cavern is almost constant in height (1300 m) but there is variation in the height along the higher latitude (9.97-9.98 deg) and this 
    variation is around 50 m. We estimated the uncertainties in surface muon energy due to the variation in height and this error is incorporated in muon charge ratio 
    results, as shown in Figure 13. 
 
 \section{Discussion}
   This work describes the method used to simulate the muon flux and charge ratio at the underground INO-ICAL detector. Using the real topography 
   of the rock overburden, we have estimated the threshold energy of the muon to reach the detector from the top surface of INO rock overburden.   
   Our work in this paper is divided into three parts: in the first part muon charge ratio is estimated using CORSIKA flux, in the second 
   part muon charge ratio is estimated using the "pika" model and finally muon charge ratio results estimated by CORSIKA are compared with the results obtained by "pika" 
   model and available experimental data. The analysis is performed only for the downgoing atmospheric muons. The muon charge ratio analysis at INO-ICAL using 
   CORSIKA flux is limited to the flux entering from the top surface of the detector only. The same analysis is repeated using 
   the "pika" model which incorporates five out of six surfaces of the detector. The muon charge ratio estimation is based on the strength of the 
   detector's magnetic field.  
   
   \begin{figure}
\centering
\includegraphics[scale=0.5]{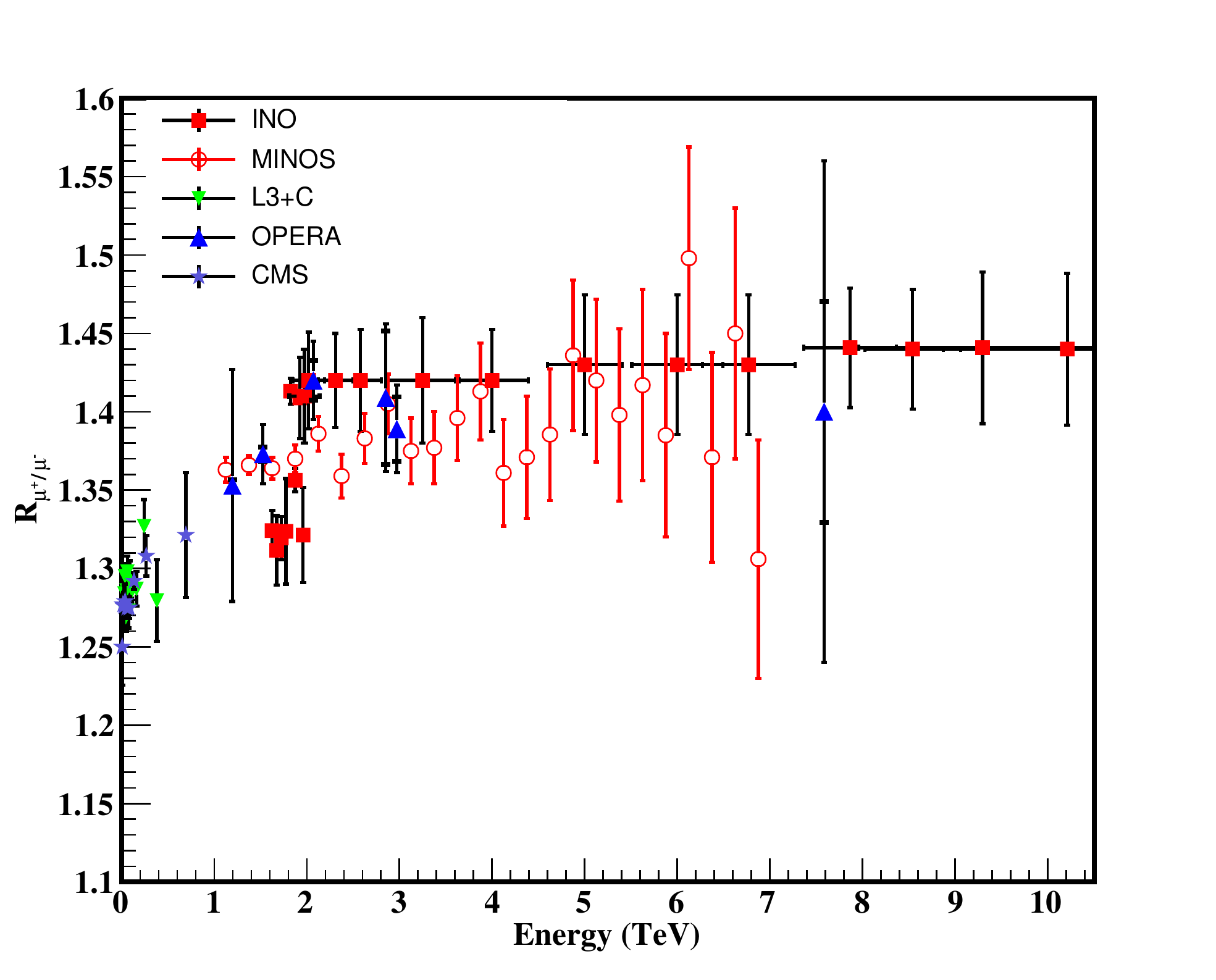}
\caption{\label{fig:INO_exp_ratio2}Muon charge ratio for a wide range of zenith angles($\theta$=$0^{0}-60^{0}$) at INO as a function of surface muon energy and 
their comparison with standard existing experimental results \cite{2,3,4,5,6}. The charge ratio results estimated from CORSIKA are also shown in this Figure (red blocks) 
till 2 TeV lying in the ratio range around 1.3 to 1.4.}
\end{figure}

   Due to its large dimensions, ICAL detector can detect muons coming from large zenith angles, as discussed in the section 3(3.2 $\&$ 3.3). Momentum 
 reconstruction and charge identification efficiency is more than 90$\%$ in 1-100 GeV energy range but falls with increasing energy. The ICAL 
 charge identification efficiency drops from 80$\%$ to 70$\%$ for the input momentum range 200 GeV/c to 400 GeV/c. Our analysis done with CORSIKA flux is limited 
 to momentum reconstruction and charge identification efficiency of the detector more than 70$\%$, as shown in Fig. 4 $\&$ 7. The analysis of detector efficiency is 
 performed at various fixed energies and zenith angles while the azimuthal angle is uniformly averaged over the entire range -$\pi\leq\phi\leq\pi$. A detailed 
 simulation is peformed with real topography of the surface surrounding the INO location for the zenith angle range $0-60^{0}$ and azimuthal angle 
 -$\pi\leq\phi\leq\pi$.

      Estimation of muon charge ratio at the underground INO-ICAL detector as well as at the top of the hill surface is performed using CORSIKA flux 
      having surface muon energy in the range 1.60-2.00 TeV. The estimated muon charge ratio varies between 1.3-1.4 as shown in Figure 13. The 
      average value of the charge ratio observed at underground detector is 1.347$\pm$0.019 and corresponding average muon charge ratio at the 
      surface is 1.358. Charge ratio observed in the same energy range by L3+C\cite{2}, MINOS\cite{3}, CMS\cite{4} and OPERA\cite{5} experiments are shown in 
      fig. 14. 

      The muon charge ratio analysis, at higher energies and higher zenith angles is included in our work by using "pika" model. For the estimation 
      of the muon charge ratio, the muon momentum reconstruction efficiency and charge identification efficiency of INO-ICAL detector are taken 
      into consideration. And charge ratio for surface muon energy in the range 1.6-10.0 TeV surface muon energy is estimated. The estimated muon 
      charge ratio as a function of $E_{\mu}cos\theta$ is shown in Fig. 13. The estimated average muon charge ratio is found to be 1.437$\pm$ 0.055, 
      which is slightly higher than the muon charge ratio results estimated by available experimental data.

\section{Conclusion}
            In this work we are presenting the cosmic rays physics potential of INO-ICAL and demonstrate that INO-ICAL will be able to give us more and more useful 
            information with respect to other past experiment having the possibility to really measure the muon charge ratio in a much larger energy range. 
            The estimated results using CORSIKA can be used for relating the secondary spectrum corresponding to the primary cosmic rays spectrum, which will be 
            helpful for studying the primary cosmic rays spectrum. Charge ratio estimation is further extended for larger zentith angle by using the analytical model 
            where the muon's will travel through more rock cover and this will enable us to go to probe charge ratio at much higher energy range. Muon charge ratio 
            estimation using both the approach for vertical as well as larger zenith angle can be used for estimating the neutrino flux in the lower and higher energy 
            range at the detector.  
 
  \section{Acknowledgment} 
         
         This work is partially supported by Department of Physics, Lucknow University, Department of Atomic Energy, Harish-Chandra Research Institute, Allahabad and 
         INO collaboration. Financially it is supported by Government of India , DST Project no-SR/MF/PS02/2013, Department of Physics, Lucknow University. We thank 
         Prof. Raj Gandhi for useful discussion, providing the cluster facility for data generation and accomodation to work in HRI. Real hill topography is provided 
         by IMSC group so we thank them all for giving this useful information and the discussion specially with Meghna, Lakshmi, Sumanta, Prof. D. Indumathi and 
         Prof. M.V.N. Murthy. We thank alot to Prof. G. Majumder for, code-related discussion and implementation and many clarifications.

\end{document}